\documentstyle[twoside,fleqn,espcrc2,epsf]{article}

\makeatletter
\def\lesssim{\mathrel{\mathpalette\vereq<}}
\def\vereq#1#2{\lower3pt\vbox{\baselineskip1.5pt \lineskip1.5pt
\ialign{$\m@th#1\hfill##\hfil$\crcr#2\crcr\sim\crcr}}}

\def\alt{\lesssim}

\makeatother

\begin{document}

\title{
{\normalsize E-print hep-ph/9812408 \hfill Preprint YARU-HE-98/08}\\[4mm] 
{\bf Field-induced axion decay $a \to e^+ e^-$ via plasmon}}

\author{N.V.~Mikheev$^{\rm a}$, A.Ya.~Parkhomenko\address{Yaroslavl State 
       (Demidov) University, Sovietskaya 14, Yaroslavl 150000, Russia}, 
        and 
        L.A.~Vassilevskaya$^{{\rm a},}$\address{Moscow State (Lomonosov) 
        University, V-952, Moscow 117234, Russia}}


\begin{abstract}
The axion decay $a \to e^+ e^-$ via a plasmon is investigated 
in an external magnetic field. The results we have 
obtained demonstrate a strong catalyzing influence of the field
as the axion lifetime in the magnetic field of order $10^{15}$~G
and at temperature of order 10~MeV is reduced to $10^{4}$~sec.

\end{abstract}


\maketitle

\thispagestyle{empty}


\section{\uppercase{Introduction}}

The pseudo-Goldstone boson associated with Peccei-Quinn symmetry 
$U_{PQ}(1)$~\cite{Peccei77}, the axion~\cite{WW}, is of interest not only 
in theoretical aspects of elementary particle physics, but in some 
astrophysical and cosmological ap\-pli\-ca\-tions as 
well~\cite{Turner,Raffelt90,Raffelt-book}.
It is also known that astrophysical and cosmological considerations 
leave a narrow window on the axion mass~\cite{Raffelt-castle}
\footnote{However in paper~\cite{Rubakov} 
a possibility to solve the CP problem of QCD within a GUT model 
with a heavy axion $m_a \alt 1$ TeV is considered.}:
\begin{equation}
10^{-5}~{\rm eV} \lesssim m_a \lesssim 10^{-2}~{\rm eV},
\label{eq:MA}
\end{equation}
\noindent where axions could exist and provide a significant fraction
or all of the cosmic dark matter. 

At present the interest in axions as a possible dark matter candidate
stimulates fullscale searches for galactic axions in experiments
~\cite{Kyoto,Livermore}. Negative results of experiments are
naturally explained by the fact that axions are very weakly coupling
and very long living. The axion lifetime in vacuum is gigantic one:
\begin{equation}
\tau \sim 6.3 \cdot 10^{42} \, \mbox{s} \,
\left ( {10^{-2}\mbox{eV} \over m_a } \right )^6 \;
\left ( {E_a \over 1 \mbox{MeV} } \right ) .
\label{eq:T0} 
\end{equation}

On the other hand, in some astrophysical considerations where axions
effects could be substantial, it is important to take into account 
the influence of plasma and the magnetic fields.
One of the most physically realistic situations presented in 
many astrophysical objects is that when from both these 
components of the active medium the plasma dominates.
For the physical circumstances of interest to us, the temperature $T$
appears to be the largest physical parameter. So, we will use a
well describing by a crossed field limit (${\bf E} \perp {\bf B}$, $E = B$)
case, $T^2 \gg e B \gg m_e^2$, when a great number of the Landau levels
are excited. At the same time the condition $T^2 \gg e B $ is fulfilled,
the magnetic filed is strong enough, $e B \gg m_e^2$, in comparison with
the known Schwinger value $B_e = m^2_e/e \simeq 4.41 \cdot 10^{13}$~G. 
Possible mechanisms of a generation of such strong fields 
as $B \sim 10^{15} - 10^{17}$~G in astrophysics were discussed in a
number of papers~\cite{magnetar,toroidal}.

In this paper we investigate the influence of the magnetic field and 
plasma on the axion decay into electron-positron pair via a photon 
intermediate state $a \to \gamma \to e^+ e^-$ in KSVZ model~\cite{KSVZ} 
in which axions have not direct coupling to leptons. 
The reason for which this forbidden in vacuum and plasma channel is opened 
in the magnetic field is that $e^+e^-$ pair can have both 
time-like and space-like total momentum as it occurs in photon splitting 
$\gamma \to e^+ e^-$~\cite{Klepikov}.


\section{\uppercase{Matrix Element}}

\begin{figure}[tb]
\centerline{\epsfxsize=.38\textwidth \epsffile[145 585 345 685]{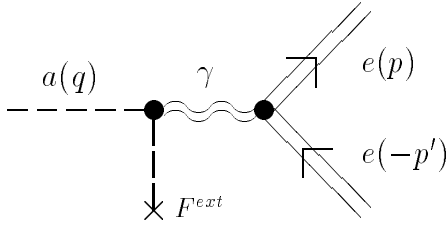}}
\caption{The diagram describing the axion decay into electron-positron
         pair via virtual photon.}
\label{fig:agee}
\end{figure}

A diagram describing $a \to e^+ e^-$ decay via the plasmon intermediate 
state is shown in Fig.~\ref{fig:agee}, 
%
%
%
%
where solid double lines imply the influence of the magnetic field 
in the electron wave functions and undulating double lines imply the 
influence of medium in the photon propagator.

The Lagrangian describing the axion-photon coupling can be presented
in the form:
\begin{eqnarray}
{\cal L}_{a \gamma}= g_{a\gamma} \,\partial_\mu A_\nu\,
\tilde F_{\nu\mu}\,a\,,
\label{eq:Lag} 
\end{eqnarray}
\noindent where $A_\mu$ is the four potential of the quantized 
electromagnetic field, $\tilde F$ is the dual external field 
tensor, $a$ is the axion field.
Here $g_{a\gamma}$ is the known axion-photon coupling 
constant with the dimension (energy)$^{-1}$~\cite{Raffelt90} 
$g_{a\gamma}=\alpha\xi/2\pi f_a$ where $\xi$ is a
model-dependent parameter, $f_a$ the Peccei-Quinn scale. 

The matrix element of $a \to e^+ e^-$ decay corresponding to the
diagram of Fig.~\ref{fig:agee} can be written as
\begin{equation}
S=\frac{g_{a\gamma}}{\sqrt{2 E_a V}} \, h J
\label{eq:S1}
\end{equation}
\noindent 
in terms of the currents
\begin{eqnarray}
J_\alpha & = & \int d^4 x\,
{\bar\psi(p,x)}\,\gamma_{\alpha}\,\psi(-p',x) \, e^{-iqx},
\nonumber \\
h_{\alpha} & = &
- i e (q \tilde F G (q))_\alpha =
- i e q_\mu \tilde F_{\mu\nu} G_{\nu\alpha}(q).
\nonumber 
\end{eqnarray}
\noindent 
Here, $e>0$ is the elementary charge, $p=(E,{\bf p})$ and
$p'=(E',{\bf p}')$ are the quasi-mo\-men\-ta of final electron 
and positron ($p^2={p'}^2=m_e^2$) in an external field; 
$q = (E_a,{\bf q})$ is the axion momentum; 
$\psi(p,x)$ is the exact solution of the Dirac equation in the 
magnetic field. The condition of the relative weakness of the
magnetic field, $e B \ll T^2$, means that the plasma influence
determines basically the properties of the photon propagator 
$G_{\alpha \beta}$ which can be presented as a sum of transverse 
and longitudinal parts:
\begin{eqnarray}
G_{\alpha \beta} & = & - i \left ( 
  \frac{ {\cal P}^{(T)}_{\alpha\beta}}{q^2 - \Pi^{(T)}} 
+ \frac{ {\cal P}^{(L)}_{\alpha\beta}}{q^2 - \Pi^{(L)}} 
\right ),
\label{eq:G} \\
{\cal P}^{(T)}_{\alpha\beta} & = & - \sum^2_{\lambda = 1} 
t_{\alpha}^{\lambda}\;t_{\beta}^{\lambda},
\nonumber \\
{\cal P}^{(L)}_{\alpha\beta} & = & - \, l_{\alpha}\;l_{\beta},
\nonumber
\end{eqnarray}
\noindent 
Here, $\Pi^{(T)}$ and $\Pi^{(L)}$ are the transverse and longitudinal 
eigenvalues of the polarization operator;
$t_{\alpha}^{\lambda} = (0, \bf t^{\lambda})$ and $l_{\alpha}$ 
denote transverse and longitudinal photon polarization vectors:
\begin{eqnarray}
{\bf t}^{(1)} & = & \frac{\bf q \times \bf B}{|{\bf q} | B \sin \theta}, \quad
{\bf t}^{(2)} = \frac{{\bf q} \times {\bf t}^{(1)}}{|\bf q |},
\nonumber \\
\ell_\alpha & = & \sqrt{\frac{q^2}{(uq)^2 - q^2}}\,
\left(u_\alpha - \frac{uq}{q^2}\,q_\alpha \right),
\nonumber
\end{eqnarray}
\noindent where $u_{\alpha}$ the four-velocity of the medium; $\theta$  
is the angle between the external magnetic field ${\bf B}$ and the 
axion momentum ${\bf q}$. 

Being integrated over the variable $x$ the expression~(\ref{eq:S1})
can be presented in the form:
\begin{eqnarray}
S & = & {(2 \pi)^4 \delta^2({\bf Q}_{\perp}) \; \delta(k Q)
\over \sqrt {2 E_a V \cdot 2 E V \cdot 2 E' V}}\; 
\frac{g_{a\gamma}}{\pi (4 \beta)^{1/3}} 
\label{eq:S2} \\
& \times & \bar U(p) \; \bigg [ \; \hat h \;\Phi (\eta) 
+ \frac{i e \ae_{-}}{2 z m_e^2} \; (\gamma F h) \; \Phi' (\eta) 
\nonumber \\ 
& - & \frac{ e \ae_{+}}{2 z m_e^2} \; \gamma_5 \; 
(\gamma \tilde F h) \; \Phi' (\eta) 
\nonumber \\
& + & \frac{m_e^2}{2 z^2}\; \frac{\hat k (kh)}{(k p)(k p')}\; 
\eta \; \Phi (\eta) \bigg ] \; U (- p'),
\nonumber \\
\ae_\pm & = & \frac{1}{\chi} \pm \frac{1}{\chi'},
\nonumber \\
z & = & \left (\frac{\chi_a}{2 \chi \chi'} \right )^{1/3},
\nonumber \\
\beta & = & \frac{1}{4} u^3 z^3, \quad  u^2 = - \frac{e^2 a^2}{m_e^2},
\nonumber \\
\chi^2 & = & \frac{e^2 (p F F p)}{m_e^2}, \quad 
{\chi'}^2 = \frac{e^2 (p' F F p')}{m_e^2}, 
\nonumber \\
\chi_a^2 & = & \frac{e^2 (q F F q)}{m_e^2}, 
\nonumber
\end{eqnarray}
\noindent 
where $Q = q - p - p'$, 
${\bf Q}_\perp$ is the perpendicular to $\bf k$ component 
(${\bf Q}_\perp {\bf k}= 0$). With the four-potential 
$A_\mu = (kx) a_\mu$ the external field
tensor is $F_{\mu\nu} = k_\mu a_\nu - k_\nu a_\mu$.
Finally, $\Phi (\eta)$ is the Airy function:
\begin{eqnarray}
\Phi (\eta) & = &  \int\limits_0^\infty d t \cos \left ( \eta t +
{t^3\over 3} \right ),  
\label{eq:Ai} \\ 
\eta & = & z^2 \; (1 + \tau^2),
\qquad
\tau = - \, \frac{e (p \tilde F q)}{m_e^4 \chi_a},
\nonumber 
\end{eqnarray}
\noindent and $\Phi' (\eta) = \partial \Phi (\eta)/ \partial \eta$.


\section{\uppercase{Decay probability}}

To obtain the decay probability one has to carry out a non-trivial 
integration over the phase space of the $e^+ e^-$ pair taking their 
specific kinematics in the magnetic field into account.

As the analysis shows the contribution of the transverse photon mode
to $a \to  e^+ e^-$ decay probability in the ultrarelativistic 
case is negligibly small. The main contribution due to the longitudinal 
plasmon intermediate state has a form:
\begin{eqnarray}
W & = &\frac{g_{a\gamma}^2 (e B)^2}{36 \pi} \, 
    \frac{E_a^3 \, \cos^2 \theta }
    {(E_a^2 - {\cal E}^2)^2 + \gamma^2 {\cal E}^4} \, \rho,
\label{eq:prob} \\
\rho & = & 6 \int\limits_0^1 dx \, x (1 - x) \, (1 - n) \, (1 - \bar n) ,  
\nonumber \\
n & = & \left ( \exp \frac{x E_a - \mu}{T} + 1 \right )^{-1} , 
\nonumber \\
\bar n & = & \left ( \exp \frac{(1 - x) E_a + \mu}{T} + 1 \right )^{-1} , 
\nonumber 
\end{eqnarray}
\noindent 
where $n$ and $\bar n$ are the Fermi-Dirac distributions of electrons 
and positrons at a temperature $T$ and a chemical potential $\mu$, 
respectively.
The function $\rho (E_a, T, \mu)$ has a meaning of the average value of 
supressing statistical factors and is, in general case, inside the 
interval $0 < \rho < 1$. 

\begin{figure}[tb]
\centerline{\epsfxsize=.38\textwidth \epsffile[115 430 395 710]{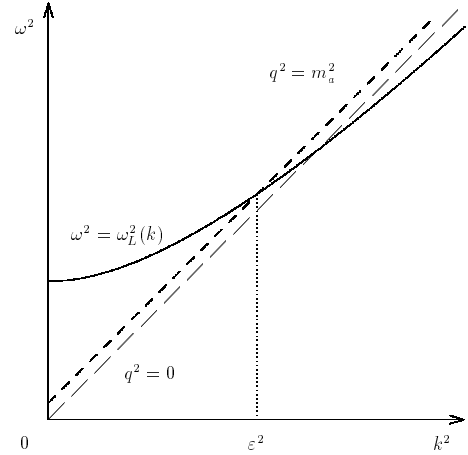}}
\caption{Dispersion relations $\omega^2=\omega^2_L(k)$ for
         longitudinal plasmons (solid line), axions 
         $E_a^2 = k^2+m^2_a$ (short dashes), and vacuum photons 
         $\omega=k$ (long dashes).}
\label{fig:disp}
\end{figure}

Eq.~(\ref{eq:prob}) has a resonant behaviour at the point 
$(E_a^2)_{res} = {\cal E}^2$ where axion and longitudinal 
plasmon dispersion curves cross (Fig.~\ref{fig:disp}).  
%
%
%
%
The dimensionless resonance width $\gamma$ of the $a \to e^+ e^-$ 
process in Eq.~(\ref{eq:prob}) is 
\begin{equation}
\gamma=\frac{{\cal E} \Gamma_L({\cal E})}{q^2 Z_L},
\label{eq:gamma}
\end{equation}
\noindent where $\Gamma_L ({\cal E})$ is the total width of the 
longitudinal plasmon; $Z_L$ is the renormalization factor of 
longitudinal plasmon wave-function: 
\begin{equation} 
Z_L^{-1} = 1 - \frac{\partial \,\Pi^{(L)}}{\partial\,q_0^2 }. 
\label{eq:Gamma1} 
\end{equation} 
\noindent Notice that without the external field the plasmon
decay into neutrino pair takes place only. In the presense of the magnetic
field which, from one hand, is weak, $e B \ll E^2$, and, from the other
hand, strong enough, $e B \gg \alpha^3 E^2$, novel channel of the
longitudinal plasmon decay is opened $\gamma_L\to e^+e^-$. 
However the main contribution to the width $\Gamma_L ({\cal E})$ is 
determined by the process of the longitudinal 
plasmon absorbtion $\gamma_L e^- \to e^-$ which becomes possible in this
kinematical region in the magnetic field.

Below we give the expressions for ${\cal E}^2$ and $\gamma$ in two limits: 
\newline
i) degenerate plasma 
\begin{eqnarray}
{\cal E}^2 & \simeq & \frac{4 \alpha}{\pi} \, \mu^2 \, 
\left ( \ln \frac{2 \mu}{m_e} - 1 \right ) , 
\label{eq:degen} \\
\gamma & \simeq & \frac{2 \alpha}{3} \, \frac{\mu^2}{{\cal E}^2} ,
\nonumber
\end{eqnarray}
ii) nondegenerate hot plasma 
\begin{eqnarray}
{\cal E}^2 & \simeq & \frac{4 \pi \alpha}{3} \, T^2 \, 
\left ( \ln \frac{4 T}{m_e} - 0.647 \right ) , 
\label{eq:nondegen} \\
\gamma & \simeq &\frac{2 \alpha}{3} \, \frac{\mu^2}{{\cal E}^2} .
\nonumber 
\end{eqnarray}
Considering possible applications of the result we have obtained to 
cosmology it is necessary to take an influence of a hot plasma into 
account. Under the early Universe conditions the hot plasma is 
nondegenerate one ($\mu \ll T$) and the medium parameter 
$\rho$ is inside the interval $1/4 < \rho < 1$.
With ${\cal E}^2$ and $\gamma$ from~(\ref{eq:nondegen}) 
we obtain the following estimation for  
the axion lifetime in the resonance region:
\begin{eqnarray}
& & \tau (a \to \gamma_{pl} \to e^+ e^-) \simeq 
 2.5 \cdot 10^4 \, \mbox{s} \, 
\label{eq:time-KSVZ} \\ 
& & \times 
\left ( \frac{10^{-10}}{ g_{a \gamma} \, \mbox{GeV}} \right )^2 \, 
\left ( \frac{T}{10 \, \mbox{MeV}} \right ) \,
\left ( \frac{10^{15} \, \mbox{G}}{B} \right )^2.
\nonumber 
\end{eqnarray}
It is interesting to compare~(\ref{eq:time-KSVZ}) with the field-induced 
axion lifetime~\cite{Ljuba-aff97} in the model~\cite{DFSZ}
where axions couple with electrons on the tree level:
\begin{eqnarray}
& & \tau (a \to e^+ e^-) \simeq 3.4 \cdot 10^6 \,\mbox{s} 
\label{eq:time-DFSZ} \\
& & \times 
\left ( \frac{10^{-13}}{g_{a e}} \right )^2 \, 
\left ( \frac{T}{10 \, \mbox{MeV}} \right )^{1/3} \,
\left ( \frac{10^{15} \, \mbox{G}}{B} \right )^{2/3}.
\nonumber 
\end{eqnarray}
The expressions~(\ref{eq:time-KSVZ}) and~(\ref{eq:time-DFSZ}) we
have presented here demonstrate the strong catalyzing influence of
the medium, plasma and the magnetic field, on the axion lifetime in 
comparison with the vacuum one~(\ref{eq:T0}).
Due to the resonance behaviour of $a \to \gamma_{pl} \to e^+ e^-$
via the longitudinal plasmon the axion lifetime in KSVZ model with
induced axion-electron interaction can be smaller than in 
DFSZ model with the direct coupling.

\section*{\uppercase{Acknowledgements}}

N.~Mikheev and L.~Vassilevskaya thank the organizers of the 5-th IFT 
Workshop on Axions for their warm hospitality during the visit.
This research was partially supported by INTAS under grant No.~96-0659
and by the Russian Foundation for Basic Research under grant 
No.~98-02-16694. 


\end{document}